\documentclass[prl,twocolumn,preprintnumbers,superscriptaddress,showpacs]{revtex4-1}
\usepackage[colorlinks=true, pdfstartview=FitV, linkcolor=red, citecolor=blue, urlcolor=black, pdftitle={},pdfauthor={Tomoya Hayata},pdfsubject={}, pdfkeywords={}]{hyperref}

\usepackage{graphicx}
\usepackage{dcolumn}
\usepackage{bm}
\usepackage{subfigure}
\usepackage{slashed}
\usepackage{comment}
\usepackage{booktabs}
\usepackage{amssymb}
\usepackage{amsmath}
\usepackage{color}

\begin{document}

\title{Relativistic Chiral Kinetic Theory from Quantum Field Theories}
\preprint{RIKEN-QHP-257, RIKEN-STAMP-31}
\author{Yoshimasa Hidaka}
\affiliation{
Theoretical Research Division, Nishina Center,
RIKEN, Wako, Saitama 351-0198, Japan} %
\author{Shi Pu}
\affiliation{
Department of Physics, The University of Tokyo,
7-3-1 Hongo, Bunkyo-ku, Tokyo 113-0033, Japan} %
\author{Di-Lun Yang}
\affiliation{
Theoretical Research Division, Nishina Center,
RIKEN, Wako, Saitama 351-0198, Japan} %
\date{\today}
\begin{abstract}
The chiral kinetic theory of Weyl fermions with collisions in the presence of weak electric and magnetic fields is derived from quantum field theories. It is found that the side-jump terms in the perturbative solution of Wigner functions play a significant role for the derivation. Moreover, such terms manifest the breaking of Lorentz symmetry for distribution functions. The Lorentz covariance of Wigner functions thus leads to modified Lorentz transformation associated with side-jump phenomena further influenced by background fields and collisions.
\end{abstract}

\maketitle

\textit{Introduction}.---
Novel quantum transport processes in Weyl fermionic systems have been widely investigated, in particular for the so-called chiral magnetic and vortical effects  such that
charged currents are induced by magnetic and vortical
fields \cite{Kharzeev:2007jp,Fukushima:2008xe,Kharzeev:2009pj}. These effects associated with quantum anomaly have been studied from different theoretical approaches including 
relativistic hydrodynamics \cite{Son:2009tf,Sadofyev:2010pr,Pu:2010as,Kharzeev:2011ds},
lattice simulations \cite{Abramczyk:2009gb,Buividovich:2009wi,Buividovich:2009zzb,Buividovich:2010tn,Yamamoto:2011gk}, and
gauge/gravity duality \cite{Erdmenger2009,Torabian2009a,Banerjee2011}. These effects might be (in-)directly observed in heavy ion collisions \cite{Kharzeev:2015znc} and in condensed matter systems such as Weyl semimetals \cite{Li:2014bha}.

From both theoretical and experimental perspectives, it is imperative to understand these anomalous effects in non-equilibrium conditions. One promising approach is kinetic theory, which can delineate
non-equilibrium transport of a particle when the interaction and background fields are sufficiently weak. Nevertheless, it is hard to incorporate anomalous effects through the standard Boltzmann equations \cite{Pu:2010as}. The chiral kinetic theory (CKT), which describes anomalous transport of Weyl fermions, has been thus developed from the path-integral \cite{Stephanov:2012ki}, Hamiltonian \cite{Son:2012wh}, and local-equilibrium quantum kinetic approaches~\cite{Gao:2012ix,Chen:2012ca}. In such formalism, the effective velocity and forces for a single particle are modified by the Berry curvature $\boldsymbol{\Omega}_\mathbf{p}=\mathbf{p}/(2|\mathbf{p}|^{3})$,
where $\mathbf{p}$ represents the spatial momentum of the particle, which originates from the Berry phase in an adiabatic process \cite{Berry1984}. Further generalization to massive Dirac fermions can be found in Ref.~\cite{Chen:2013iga}. In order to bridge the semi-classical approaches~\cite{Stephanov:2012ki,Son:2012wh} and quantum field theories, the CKT is also derived from Wigner functions in the high-density effective theory~\cite{Son:2012zy} (see also Ref.~\cite{Manuel:2014dza} for relevant study of the on-shell effective field theory.)

However, there still exist potential issues in the chiral kinetic equation. First, the field-theory derivation in Refs.~\cite{Son:2012zy} and \cite{Gao:2012ix,Chen:2012ca} are subject to a predominant chemical potential and local equilibrium, respectively. The derivation for more general systems beyond local equilibrium is thus needed. Second, the non-manifestation of Lorentz invariance in the chiral kinetic equation has been recently discussed in Refs.~\cite{Chen:2014cla,Chen:2015gta} from the {semi-classical} approach. The authors {propose} that the ordinary Lorentz transformation is modified by ``side-jump'' phenomena necessary to ensure angular-momentum conservation in collisions, while the same scenario is not fully understood in quantum field theories. Furthermore, it is more systematic to incorporate collisions in the field-theory framework.


In this letter, we address the aforementioned issues related to Weyl fermions in quantum field theories. It turns out that these issues are in fact connected. By solving Dirac equations, nontrivial side-jump terms coupled to background fields and self-energy from collisions naturally appear in the perturbative solution for Wigner functions up to $\mathcal{O}(\hbar)$ in terms of $\hbar$ expansion (equivalent to the gradient expansion as the long-wavelength approximation), which allude to modified Lorentz transformation of coordinates and momenta for distribution functions. The findings based on field theories support the modified Lorentz transformation proposed in \cite{Chen:2014cla,Chen:2015gta} and further incorporate the effects from background fields and collisions simultaneously. For free fermions, we further show that the side-jumps are related to reparametrization of distribution functions. From field theories, a self-consistent expression of the CKT with collisions and background fields is systematically derived.   

\textit{Side-jumps from Wigner functions}.--- We start from Dirac equations for non-interacting right-handed Weyl fermions,        
\begin{eqnarray}
\slashed{D}_xS^{<(>)}(x,y)=S^{<(>)}(x,y)\slashed{D}_y^{\dag}=0,
\end{eqnarray}
where $\slashed{D}=\sigma^{\mu}D_{\mu}$ with $D_{\mu}=(\partial_{\mu}+iA_{\mu}/\hbar)$ and $\sigma^{\mu}=(1,\boldsymbol{\sigma})$ for $\sigma^i$ being Pauli matrices. 
Here $S^<(x,y)=\langle\psi^{\dagger}(y)\psi(x)\rangle$ and $S^>(x,y)=\langle\psi(x)\psi^{\dagger}(y)\rangle$ correspond to lesser and greater propagators in position space and hereafter we focus on just $S^<(x,y)$.  In this Letter, we include the electric charge into the definition of the gauge field.
We take the mostly negative Minkowski metric for convention. By taking $Y=x-y$ and $X=(x+y)/2$, we carry out the Wigner transformation 
\begin{eqnarray}
\grave{S}^<(q,X)=\int d^4Ye^{\frac{iq\cdot Y}{\hbar}}S^<(x,y),
\end{eqnarray}
where the gauge link is implicitly embedded and $q_{\mu}$ denotes the canonical momentum. The Dirac equations after Wigner transformation up to $\mathcal{O}(\hbar)$ become
\begin{eqnarray}\label{linear_Dirac}
\sigma^{\mu}\left(\hbar\Delta_{\mu}-2iq_{\mu}\right)\grave{S}^<=0, \
\left(\hbar\Delta_{\mu}+2iq_{\mu}\right)\grave{S}^<\sigma^{\mu}=0,
\end{eqnarray} 
where $\Delta_{\mu}=\partial_{\mu}+F_{\nu\mu}{\partial}/{\partial q_{\nu}}$ and $\partial_{\mu}=\partial/\partial X^{\mu}$ \cite{Elze:1986qd}.
Adding and subtracting two equations above give the {\it difference equations}: 
\begin{eqnarray}\label{diff_Dirac}\nonumber
&&\hbar\left\{\sigma^{\mu},\Delta_{\mu}\grave{S}^<\right\}-2i\left[q_{\mu}\sigma^{\mu},\grave{S}^<\right]=0,
\\
&&\hbar\left[\sigma^{\mu},\Delta_{\mu}\grave{S}^<\right]-2i\left\{q_{\mu}\sigma^{\mu},\grave{S}^<\right\}=0,
\end{eqnarray}
where $\left\{A,B\right\}=AB+BA$ and $\left[A,B\right]=AB-BA$. Note that these equations are up to all orders of $\hbar$ for constant background fields. 

The task now is to solve Eq.~(\ref{diff_Dirac}) perturbatively including quantum corrections up to $\mathcal{O}(\hbar)$. We may take
\begin{eqnarray}
\grave{S}^<=\bar{\sigma}^{\mu}\grave{S}^<_{\mu},\quad \grave{S}^<_{\mu}=\grave{S}^{(0)<}_{\mu}+\hbar\delta\grave{S}^<_{\mu},
\end{eqnarray}
where $\bar{\sigma}^{\mu}=(1,-\boldsymbol{\sigma})$. It is trivially to show that the leading-order solution reads $\grave{S}^{(0)}_{\mu}=(2\pi)q_{\mu}\delta(q^2)f$ and $q^{\mu}\Delta_{\mu}f=0$ up to $\mathcal{O}(1)$, where $f(q,X)$ denotes the distribution function. For simplicity, we first consider the case that $f=1$ as a constant. The first equation in Eq.~(\ref{diff_Dirac}) then results in 
\begin{eqnarray}\label{solve_Sc}
(F^{ij}q_0q_j-F^{0i}|{\bf q}|^2+F_{j 0}q^iq^j)\frac{\partial\delta(q^2)}
{\partial q^2}=\epsilon^{ijk}q_j\frac{\delta\grave{S}^{<}_{k}}{2\pi},
\end{eqnarray}
where we apply $q^2\partial\delta(q^2)/\partial q^2=-\delta(q^2)$ in the derivation.
From Eq.~(\ref{solve_Sc}), one finds the $\mathcal{O}(\hbar)$ correction for constant $f$ takes the form,
\begin{eqnarray}\label{trace_sol}
\grave{S}^{c<}_{\mu}=(2\pi)\left(q_{\mu}\delta(q^2)+\hbar \epsilon_{\mu\nu\alpha\beta}q^{\nu}F^{\alpha\beta}\frac{\partial \delta(q^2)}{2\partial q^2}\right)f, 
\end{eqnarray} 
where $\epsilon^{\mu\nu\alpha\beta}$ represents the Levi-Civita tensor with $\epsilon^{0123}=\epsilon^{123}=1$. The superscript $c$ denotes that the spectral density (the part aside from $f$) is Lorentz covariant.  
Nevertheless, when $f$ is non-constant, this is not the complete solution.

Coming back to the Dirac equations in Eq.~(\ref{diff_Dirac}), it is then more convenient to separate the trace and traceless part (linear to $\sigma^i$), which yields
\begin{eqnarray}\label{trace_part}
\Delta_{\mu}\grave{S}^{\mu <}=0,\quad q^{\mu}\grave{S}_{\mu}^<=0,
\end{eqnarray}
and
\begin{eqnarray}\label{traceless_part}
\hbar\Delta_{[i}\grave{S}^<_{0]}-2\epsilon^{ijk}q_j \grave{S}^<_k=0,  \
\hbar\epsilon^{ijk}\Delta_j\grave{S}_k^<+2q_{[i}\grave{S}^<_{0]}=0.
\end{eqnarray}  
Now the perturbative solution is actually solved from the traceless part in Eq.~(\ref{traceless_part}), while $q^{\mu}\grave{S}^<_{\mu}=0$ gives the constraint and $\Delta_{\mu}\grave{S}^{<\mu}=0$ yields the kinetic theory.       
It turns out that Eq.~(\ref{traceless_part}) results in the side-jump term on top of Eq.~(\ref{trace_sol}). From Eq.~(\ref{traceless_part}), one finds \footnote{Note that one needs to use $\delta(q^2)q^{\mu}\Delta_{\mu}f=0$ to substitute $\Delta_0f$ with $q_0^{-1}{\bf q}\cdot{\bf \Delta}f$ in order to show the solution satisfies the first equation in Eq.~(\ref{traceless_part}). Also, one can show that Eq.~(\ref{trace_sol}) satisfies Eqs.~(\ref{trace_part}) and (\ref{traceless_part}) for constant $f$.}
\begin{eqnarray}\label{full_sol}
\grave{S}^<_{\mu}&&=\grave{S}^{c<}_{\mu}+\hbar\delta\grave{S}^{f<}_{\mu},\\
\delta\grave{S}^{f<}_{\mu}&&=2\pi\delta_{\mu i}\epsilon_{ijk}\frac{\delta(q^2)q_j}{2q_0}\Delta_{k}f,
\end{eqnarray}
where the superscript $f$ here denotes that the spectral density of the side-jump term $\delta\grave{S}^{f<}_{\mu}$ is not Lorentz covariant (frame dependent) \footnote{Note that the gradient expansion implicitly introduces an infrared cutoff for $|{\bf q}|$, which has to be larger than the scale of gradient or background fields.}. One may recognize that $\delta\grave{S}^{f<}_{\mu}$ corresponds to the magnetic-moment coupling in the absence of background fields in Refs.~\cite{Son:2012zy,Chen:2014cla} or the vorticity-related term in local equilibrium \cite{,Gao:2012ix,Chen:2012ca}. The solution in Eq.~(\ref{full_sol}) is not unique: We may further introduce arbitrary terms proportional to $\hbar q_{\mu}\delta(q^2)$, while they could be absorbed into $f$. This degeneracy is actually pertinent to Lorentz covariance of Wigner functions, which will be discussed later.

\textit{Chiral kinetic theory}.---
By inserting the perturbative solution up to $\mathcal{O}(\hbar)$ in Eq.~(\ref{full_sol}) into Eq.~(\ref{trace_part}), $q_{\mu}\grave{S}^{<\mu}=0$ is trivially satisfied, whereas $\Delta_{\mu}\grave{S}^{<\mu}=0$ gives the CKT. In the following, we present some critical steps of the derivation. Now, $\Delta_{\mu}\grave{S}^{<\mu}=0$ leads to 
\begin{eqnarray}\nonumber
&&\left(\delta(q^2)q^{\mu}
+\hbar \epsilon^{\mu\nu\alpha\beta}q_{\nu}F_{\alpha\beta}\frac{\partial\delta(q^2)}{2\partial q^2}\right)\Delta_{\mu}f\\
&&+\hbar   \epsilon^{0\beta \nu\mu} F_{\alpha\beta}\left(\frac{\partial}{\partial q_\alpha}\frac{q_\nu \delta(q^2)}{2{q_0}}\right) \Delta_{\mu}f\notag\\
&&+\hbar \epsilon^{0\beta \nu\mu}\frac{q_\nu\delta(q^2)}{2{q_0}} (\partial_\beta F_{\alpha\mu})\frac{\partial}{\partial q_\alpha}f
=0.
\end{eqnarray}
For simplicity, we consider only the particle with positive energy. 
Performing some computations, we obtain
\begin{eqnarray}\nonumber\label{ch_kinetic_0}
&&{	\delta\left(q^2+\hbar {\frac{{\bf B}\cdot {\bf q} }{q_0}}\right)\Bigg[q^\mu\Delta_\mu
	-\frac{\hbar\partial_k({\bf B}\cdot{\bf q})}{2{q_0}}\frac{\partial}{\partial q_k} }
	\\
&&{\quad+\frac{\hbar\epsilon^{ijk}E_iq_j}{2q_0^2}\Delta_k\Bigg]f=0,}
\end{eqnarray} 
where $E^i\equiv F^{i0}$ and $B^{i}\equiv-\epsilon^{ijk}F_{jk}/2$ (equivalently $F_{ij}=\epsilon_{ijk}B^k$), and we utilize $\boldsymbol{\nabla}\cdot{\bf B}=0$. 
 
The on-shell condition $q^2+\hbar{\bf B}\cdot {\bf q}/q_0=0$ now implies the shift of energy,
\begin{eqnarray}
q_0=\epsilon_{\bf q}=|{\bf q}|\left(1-\frac{\hbar {\bf B}\cdot {\bf q}}{2|{\bf q}|^3}\right).
\end{eqnarray} 
We hereby introduce an effective velocity
\begin{eqnarray}
{\bf \tilde{v}}=\frac{\partial \epsilon_{\bf q}}{\partial {\bf q}}=\frac{{\bf q}}{|\bf q|}-\frac{\hbar {\bf B}}{2|{\bf q}|^2}+\frac{\hbar ({\bf B}\cdot{\bf q}){\bf q}}{|{\bf q}|^4}.
\end{eqnarray}
Eventually, we reproduce the CKT in Ref.~\cite{Son:2012zy},
\begin{eqnarray}\nonumber\label{CKT0}
&&\Bigg[\left(1+\hbar{\bf B}\cdot\boldsymbol{\Omega}_\mathbf{q}\right)\partial_t
+\left({\bf \tilde{v}}+\hbar{\bf E}\times \boldsymbol{\Omega}_\mathbf{q}+\hbar {\bf (\tilde{v}}\cdot\boldsymbol{\Omega}_\mathbf{q}){\bf B}\right)\cdot\boldsymbol{\nabla}
\\
&&+\left({\bf \tilde{E}}+{\bf \tilde{v}}\times{\bf B}+\hbar({\bf \tilde{E}}\cdot{\bf B})\boldsymbol{\Omega}_\mathbf{q}\right)\cdot\frac{\partial}{\partial {\bf q}}\Bigg]f=0,
\end{eqnarray}
where ${\bf \tilde{E}}={\bf E}-\boldsymbol{\nabla} \epsilon_{\bf q}$.
As a cross check, we may also evaluate the currents from the perturbative solution in Eq.~(\ref{full_sol}). The particle number density is given by
\begin{eqnarray}\label{eq:charge}
J^0=\int\frac{d^4q}{(2\pi)^4}\text{Tr}\left(\grave{S}^<\right)
=\int\frac{d^3q}{(2\pi)^3}\left(1+{\hbar{\bf B}\cdot\boldsymbol{\Omega}_{\bf q}}\right)f.\nonumber\\
\end{eqnarray}
Also, the spatial current reads
\begin{eqnarray}\label{eq:current}
&&{\bf J}=\int\frac{d^4q}{(2\pi)^4}\text{Tr}\left(\mathbf{\sigma}\grave{S}^<\right)
\\\nonumber
&&\quad=\int\frac{d^3q}{(2\pi)^3} \Bigl({\bf \tilde{v}}
+\hbar{\bf E}\times \mathbf{\Omega}_{\bf q}\notag\\
&&\qquad-\hbar\epsilon_{{\bf q}}{\bf B} \,\mathbf{\Omega}_{\bf q}\cdot\frac{\partial}{\partial {\bf q}}
-\hbar\epsilon_{{\bf q}}\mathbf{\Omega}_{\bf q}\times\mathbf{\nabla}
\Bigr)f,
\end{eqnarray}
where we take the integration by part. 
The current incorporates the anomalous Hall effect, chiral magnetic effect, and side-jump as shown in Ref.~\cite{Son:2012zy}.
We notice that the expression in our formalism is valid up to $\mathcal{O}(\hbar)$.

\textit{Lorentz invariance}.--- Although we derive the CKT and anomalous effects from the perturbative solution in Eq.~(\ref{full_sol}), 
one may be concerned about the Lorentz covariance of Wigner functions due to the side-jump term. 
Since the Wigner function should be Lorentz covariant, the existence of the side-jump term suggests that $f$ is not a Lorentz scalar~\cite{Chen:2014cla,Chen:2015gta}. To understand the modifications of Lorentz transformation on $f$, it is instructive to manifest the frame (in)dependence of the side-jump term. 
Accordingly, we may generalize Eq.~(\ref{full_sol}) as
\footnote{One can in fact consider the frame dependence of Eq.~(\ref{traceless_part}) and check that the following solution indeed satisfies both equations therein.}
\begin{eqnarray}\label{side_jump_uframe}
\delta\grave{S}^{f<}_{\mu}=2\pi\delta(q^2)\epsilon_{\mu\nu\alpha\beta}\frac{q^{\alpha}u^{\beta}}{2q\cdot u}\Delta^{\nu}f^{(u)}, 
\end{eqnarray}   
where $u^{\mu}$ is the four velocity of a frame. The solution in Eq.~(\ref{full_sol}) corresponds to $u^{\mu}=(1,{\bf 0})$. {Now, one may consider a Lorentz transformation $X'^{\mu}=\Lambda^{\mu}_{\mbox{ }\nu}X^{\nu}$ and $q'^{\mu}=\Lambda^{\mu}_{\mbox{ }\nu}q^{\nu}$, which is in fact equivalent to the transformation between frames $u'^{\mu}=(\Lambda^{-1})_{~\nu}^{\mu}u^{\nu}$.
From Eq.~(\ref{side_jump_uframe}), by taking $f^{'(u)}(q',X')=f^{(u)}(q,X)+\hbar\delta f^{(u)}(q,X)$, we find
\begin{eqnarray}\label{S_frame_diff}
&&(\Lambda^{-1})^{~\nu}_{\mu}\grave{S}'^{<}_{\nu}\left(q',X'\right)-\grave{S}^{<}_{\mu}\left(q,X\right)
\\\nonumber
&&=\hbar2\pi\delta(q^2)
\left(q_{\mu}\delta f^{(u)}
+\epsilon_{\mu\nu\alpha\beta}\left(\frac{q^{\alpha}u'^{\beta}}{2q\cdot u'}-\frac{q^{\alpha}u^{\beta}}{2q\cdot u}\right)\Delta^{\nu}f^{(u)}
\right).
\end{eqnarray}
Since Wigner functions are Lorentz covariant, we should have $(\Lambda^{-1})^{~\nu}_{\mu}\grave{S}^{'<}_{\nu}-\grave{S}^{<}_{\mu}=0$, which is equivalent to the frame-independence of currents.  
Making contraction with $u^{\mu}$, Eq.~(\ref{S_frame_diff}) gives rise to
\begin{eqnarray}
f'^{(u)}\left(q',X'\right)=f^{(u)}\left(q,X\right)+\hbar N^{\mu}_{uu'}\Delta_{\mu}f^{(u)}\left(q,X\right),
\end{eqnarray}} 
where 
\begin{eqnarray}
N^{\nu}_{uu'}=-\frac{\epsilon^{\mu\nu\alpha\beta}q_{\alpha}u'_{\beta}u_{\mu}}{2(q\cdot u')(u\cdot q)}.
\end{eqnarray}
This suggests that a particle makes following side-jumps in the phase space,
$X^{\mu}\rightarrow X^{\mu}+\hbar N_{uu'}^{\mu}$ and $q_{\mu}\rightarrow q_{\mu}+\hbar N_{uu'}^{\nu}F_{\mu\nu}$,
under the Lorentz transformation. These side-jumps take the same form as those in the path-integral approach \cite{Chen:2014cla,Chen:2015gta}. 

To better understand the origin of side-jumps and frame dependence of distribution functions, we consider a free fermions with positive energy in quantum field theory. To discuss the side-jump, we consider a Lorentz transformation $\Lambda$ of the wave function for massless particles, which non-trivially transforms  with an extra phase \cite{Weinberg_vol1}: 
$|p,\lambda\rangle \rightarrow e^{-i\Phi(p,\Lambda)}|\Lambda p,\lambda\rangle$,
where $\lambda$ represents helicity. 
The Lorentz transformation of the wave function of a  
particle with positive energy thus takes the form, 
\begin{eqnarray}
v_{+}(\Lambda p)=e^{i{\Phi(p,\Lambda)}}U(\Lambda)v_{+}(p),
\end{eqnarray}
where
\begin{eqnarray}
v_+(p) = 
\begin{pmatrix}
\sqrt{|{\bf p}|{+p^3}} 
\\
\frac{{p^1+ip^2}}{\sqrt{(|{\bf p}|+{p^3})}} 
\end{pmatrix}
\end{eqnarray}
with the relativistic normalization $v^{\dagger}_{+}(p)v_{+}(p)=2|{\bf p}|$. One can accordingly write down the second quantization of a field operator as 
\begin{eqnarray}\label{fermion_op}
\psi(x)=\int\frac{d^3p}{(2\pi)^3\sqrt{2|{\bf p}|}}e^{-ip\cdot x}v_{+}(p)a_{\bf p},
\end{eqnarray}
where $a^{(\dagger)}_{\bf p}$ correspond to annihilation(creation) operators. For simplicity, we drop anti-fermions, which carry negative energy.

Considering  $G^{\mu}(p',p)=v^{\dagger}_{+}(p')\sigma^{\mu}v_{+}(p)$, which is proportional to $\grave{S}^{<\mu}$, the Lorentz transformation leads to 
\begin{eqnarray}
G^{\mu}(p',p)\rightarrow e^{i\left(\Phi(p,\Lambda)-\Phi(p',\Lambda)\right)}\Lambda^{\mu}_{\mbox{ }\nu}G^{\nu}(p',p).
\end{eqnarray} 
Therefore, $G^{\mu}(p',p)$ is not a vector. Nonetheless, the extra phase does not contribute to any physical observables.
For example, perturbation theory in thermal equilibrium. A free propagator has $p'=p$, so that the phase cancels. In
contrast, in non-equilibrium, the phase cancels with the nontrivial transformation of the distribution function, i.e.,
$N(p',p)\equiv \langle {a^{\dagger}_{\bf p'}a_{\bf p} }\rangle$
transforms as $N(p',p)\rightarrow e^{-i\left(\Phi(p,\Lambda)-\Phi(p',\Lambda)\right)}N(p',p)$, due to the Lorentz covariance of field operators.

To make $N(p', p)$ a scalar, we introduce a phase field in momentum space as $\phi(p)$ such that $\phi(p)\rightarrow {\phi'( \Lambda p)}=\phi( p)-\Phi(p,\Lambda)$ under the Lorentz transformation.
We can always introduce such a phase using gauge degrees of freedom associated with a transformation that keeps $v_{+}(p)a_{\bf p}$ invariant: $v_{+}(p)\to e^{i\phi(p)}v_{+}(p)$ and $a_{\bf p}\to e^{-i\phi(p)}a_{\bf p}$.
We hereby define the Lorentz-scalar distribution function
\begin{eqnarray}
\check{N}(p',p)\equiv e^{-i\left(\phi(p)-\phi(p')\right)}N(p',p).
\end{eqnarray}
Using Eq.~(\ref{fermion_op}) and carrying out the Wigner transformation, we find
\begin{eqnarray}\nonumber
\grave{S}^<_{\mu}(q,X)&=&(2\pi){\theta(q^0)}\delta(q^2)\Big(q_{\mu}\left(1-\hbar(\partial^{\nu}_{q}\phi-a^{\nu})\partial_{\nu}\right)
\\
&&+\hbar\delta_{\mu i}\epsilon_{ijk}\frac{q_j}{2|{\bf q}|}\partial_k
\Big)
\check{f}\left(q,X\right),
\end{eqnarray}
where  ${a^{\nu}\equiv  ic^{\dagger}_{+}(q)\partial_{q}^\nu c_{+}(q)}$
denotes the Berry connection with $c_{+}(q)=v_{+}(q)/\sqrt{2|{\bf q}|}$ from non-relativistic normalization and we define
\begin{eqnarray}
\check{f}(q,X)\equiv  \int \frac{d^3\bar{p}}{(2\pi)^3}\check{N}\left(q-\frac{\bar{p}}{2},q+\frac{\bar{p}}{2}\right)e^{-i\bar{p}\cdot X}
\end{eqnarray}
as a Lorentz scalar,  where {$\bar{p}^0={\bf \bar{p}}\cdot {\bf q}/|{\bf q}|$}.
Nonetheless, for the CKT, we apply the parametrization $f(q,X)=\check{f}\left(q_{\mu},X^{\mu}-\hbar\partial_{q}^\mu\phi(q)+\hbar a^{\mu}\right)$. The distribution function $f(q,X)$ is apparently not Lorentz invariant.


\textit{Collisions}.---In quantum field theories, we may systematically incorporate collisions (see e.g. Ref.~\cite{Blaizot:2001nr}). Based on the Dyson-Schwinger equation and taking integration along the Schwinger-Keldysh contour, the Dirac equations are given by 
\begin{eqnarray}\label{SK_eq_2}
&&\left(i\slashed{D}_x-\Sigma^{\delta}(x)\right)S^<(x,y)
\\\nonumber
&&=\int^{\infty}_{-\infty} d^4z\left(\Sigma_R(x,z)S^<(z,y)-\Sigma^<(x,z)S_A(z,y)\right),
\\\nonumber
&&S^<(x,y)\left(-i\slashed{D}_y^{\dag}-\Sigma^{\delta}(y)\right)
\\\nonumber
&&=
\int^{\infty}_{-\infty} d^4z\left(S^<(x,z)\Sigma_A(z,y)-S_R(x,z)\Sigma^<(z,y)\right),
\end{eqnarray}
where $\Sigma^{<(>)}$ represent the self-energy and $\Sigma^{\delta}$ denotes the one-particle potential and the subscripts $R/A$ correspond to retarded/advanced propagators, which are defined as $S_{R}(x,y)\equiv i\theta(x_0-y_0) S_{+}(x,y)$ and $S_{A}(x,y)\equiv -i\theta(y_0-x_0) S_{+}(x,y)$ with $S_{+}(x,y)=S^>(x,y)+S^<(x,y)$. 
In general, $\Sigma_{R/A}$ are complex, where the imaginary parts contribute to the scattering cross-section and the real parts result in renormalization of propagators. Since we only focus on the scattering process, we may drop the real parts by setting $\Sigma^{\delta}=\text{Re}[\Sigma^{R/A}]=0$ and take $\text{Re}[S^R]=0$ for the same reason.  

After making the Wigner transformation of Eq.~(\ref{SK_eq_2}), the trace of difference equations leads to
\begin{eqnarray}\label{trace_collisions}
\Delta_{\mu}\grave{S}^{<\mu}=\Sigma^<_{\mu}\grave{S}^{>\mu}-\Sigma^>_{\mu}\grave{S}^{<\mu},\quad q_{\mu}\grave{S}^{{<}\mu}=0,
\end{eqnarray}
where we set $\Sigma=\sigma^{\mu}\Sigma_{\mu}$ without the loss of generality and denote $<(>)$ explicitly since the greater propagator is also involved. On the other hand, the traceless part is given by
\begin{eqnarray}\label{traceless_collisions}
&&\hbar\Delta_{[i}\grave{S}^<_{0]}-2\epsilon^{ijk}q_j\grave{S}^<_k
=\hbar\left(\Sigma^<_{[i}\grave{S}^>_{0]}-\Sigma^>_{[i}\grave{S}^<_{0]}\right),
\\\nonumber
&&\hbar\epsilon^{ijk}\Delta_j\grave{S}^<_k+2q_{[i}\grave{S}^<_{0]}
=\hbar\epsilon^{ijk}\left(\Sigma^<_j\grave{S}^>_k-\Sigma^>_j\grave{S}^<_k\right).
\end{eqnarray}  
We may now solve for the perturbative solution. Compared to the collisioneless solution in Eq.~(\ref{full_sol}),  we find that $\grave{S}^{c<}_{\mu}$ is unchanged, whereas the side jump term becomes 
\begin{eqnarray}\label{pert_sol_collisions}
\delta\grave{S}^{f<}_{\mu}=(2\pi)\delta_{\mu i}\epsilon_{ijk}\delta(q^2)\frac{q_j}{2q_0}\left(\Delta_kf-C_k\right),
\end{eqnarray}
where $C_{\beta}[f]=\Sigma^{<}_{\beta}\bar{f}-\Sigma^{>}_{\beta}f$ with $\bar{f}$ denoting the distribution function of outgoing particles.

Since the side-jump term is altered by collisions, we investigate the modified Lorentz transformation of distribution functions. We may generalize the side-jump term as
\begin{eqnarray}
\delta\grave{S}^{f<}_{\mu}=2\pi\delta(q^2)\epsilon_{\mu\alpha\beta\nu}\frac{q^{\alpha}u^{\beta}}{2q\cdot u } \left(\Delta^{\nu}f^{(u)}-C^{\nu}[f^{(u)}]\right).
\end{eqnarray} 
The Lorentz covariance of Wigner functions yields the modified Lorentz transformation on the distribution function as  
\begin{eqnarray}\label{trans_f_coll}
{f'^{(u)}}=f^{(u)}+\hbar N^{\mu}_{uu'}\left(\Delta_{\mu}f^{(u)}
-C_{\mu}[f^{(u)}]\right).
\end{eqnarray}

From Eqs.~(\ref{trace_collisions}) and (\ref{pert_sol_collisions}), we also obtain the CKT with collisions, for $u^\mu=(1,{\bf 0})$, 
\begin{eqnarray}\nonumber
&&{\rm CKT}_0-\left(1+\hbar {\bf B}\cdot\boldsymbol{\Omega}_{\bf q}\right)C_0\\
&&+(\bf{\tilde{v}}+\hbar\bf{E}\times\boldsymbol{\Omega}_{\bf q}+\hbar(\bf{\tilde{v}}\cdot\boldsymbol{\Omega}_{\bf q} )\bf{B})
\cdot{\bf C}\notag\\
&&-\hbar\epsilon_{\bf q}\mathbf{\Omega}_{\bf q}\cdot\left(\bar{f}(\mathbf{\Delta}^>\times\mathbf{\Sigma}^<)-f(\mathbf{\Delta}^<\times\mathbf{\Sigma}^>)\right)=0,
\end{eqnarray}
where $\mathbf{\Delta}^{<(>)}=\mathbf{\Delta}+\mathbf{\Sigma}^{<(>)}$ and ${\rm CKT}_0$ corresponds to the left-hand side of the collisionless CKT in Eq.~(\ref{CKT0}).

\textit{2-2 scattering without background fields}.---
Practically, it is more useful to make further approximations of the self-energy and analyze specific collisional processes. For simplicity, here we present the leading-order 2-2 Coulomb scattering between right-handed fermions with positive energy in the absence of background fields as an example. In this particular case, we will show that the center of mass frame corresponds to a ``no-jump frame'' as proposed in Ref.~\cite{Chen:2015gta}. Now, the perturbative solution reduces to
\begin{eqnarray}\nonumber
	\tilde{S}^<_{\mu}&=&2\pi\delta(q^2)\Bigl[q_{\mu}f-\frac{\hbar}{2q\cdot u} \epsilon_{\mu\nu\alpha\beta}u^{\nu}q^{\alpha}\Big(\partial^{\beta}f
	\\
	&&+\Sigma^{>\beta}f-\Sigma^{<\beta}\bar{f}\Big)\Bigr],
\end{eqnarray}   
where we write down the frame dependence for the spectral density.

The self-energy in the leading contribution corresponding to the Coulomb scattering can be expressed as
\begin{eqnarray}\nonumber
\Sigma^<_{\mu}&=&\int_{q',k,k'}\mathcal{P}(q',k,k')\tilde{S}_{\mu}^{>}(q')\big( \tilde{S}^{<}(k)\cdot \tilde{S}^{<}(k')\big),
\end{eqnarray}
and similar for $\Sigma^>_{\mu}$ by exchanging $>$ and $<$, 
where 
\begin{eqnarray}
\mathcal{P}(q',k,k')=4e^4\left(\frac{1}{(q-k)^2}+\frac{1}{(q-k')^2}\right)^2,
\end{eqnarray}
and
\begin{eqnarray}
\int_{q',k,k'}\equiv\int \frac{d^4q'd^4k d^4k'}{(2\pi)^8}\delta^{(4)}(q+q'-k-k').
\end{eqnarray}
Although the self-energy takes a complicated form, by choosing the the center of mass frame $u_c^{\mu}\equiv(q+q')^{\mu}/\sqrt{s}=(k+k')^{\mu}/\sqrt{s}$ with $s=(q+q')^2$, one can show that the expression is considerably simplified as
 \begin{eqnarray}
 &&\Sigma^<_{\mu}
 \\\nonumber
 &&=\int_{\bf q',k,k'}\mathcal{P}(q',k,k')(k\cdot k')
 \Big[q'_{\mu}\bar{f}^{(u_c)}(q')
 \\\nonumber
 &&
 -\hbar
 \epsilon_{\mu\nu\alpha\beta}
 \frac{q^{\nu}q'^{\alpha}\partial^{\beta}\bar{f}^{(u_c)}(q')}{2q\cdot q'}\Big]f^{(u_c)}(k)f^{(u_c)}(k'),
 \end{eqnarray} 
 where
 \begin{eqnarray}
 \int_{\bf q',k,k'}=\int \frac{d^3{\bf q'}d^3{\bf k}d^3{\bf k'}}{(2\pi)^{5 }}\frac{\delta^{(4)}(q+q'-k-k')}{8E_{q'}E_kE_{k'}}.
 \end{eqnarray}
 
Since both $\tilde{S}^{>\mu}$ and $\Sigma^{<}_{\mu}$ are frame independent, we can write down the distribution functions of all scattering particles in the the center of mass frame, which yields
\begin{eqnarray}\nonumber
\tilde{S}^{>\mu}\Sigma^{<}_{\mu}&=&2\pi\delta(q^2)\int_{\bf q',k,k'}\mathcal{P}(q',k,k')(k\cdot k')(q\cdot q')
\\
&&\times\bar{f}^{(u_c)}(q)\bar{f}^{(u_c)}(q')f^{(u_c)}(k)f^{(u_c)}(k'),
\end{eqnarray}
where the $\mathcal{O}(\hbar)$ correction vanishes
and similar for $\tilde{S}^{<\mu}\Sigma^{>}_{\mu}$.
Therefore, in the the center of mass frame, the kinetic theory from the first equation in Eq.~(\ref{trace_collisions}) reduces to 
\begin{eqnarray}
\partial_{\mu}\tilde{S}^{\mu <}=2\pi\delta(q^2)q^{\mu}C_{\mu}\bigl[f^{(u_c)}\bigr],
\end{eqnarray}
where 
\begin{eqnarray}
q^{\mu}C_{\mu}\bigl[f^{(u_c)}\bigr]
&=&\frac{1}{4}\int_{\bf q',k,k'}|\mathcal{M}|^2
\\\nonumber
&&
\times\Big[\bar{f}^{(u_c)}(q)\bar{f}^{(u_c)}(q')f^{(u_c)}(k)f^{(u_c)}(k')
\\\nonumber
&&-f^{(u_c)}(q)f^{(u_c)}(q')\bar{f}^{(u_c)}(k)\bar{f}^{(u_c)}(k')
\Big]
\end{eqnarray}
with  $|\mathcal{M}|^2=4\mathcal{P}(q',k,k')(k\cdot k')(q'\cdot q)$
comprises ``no side-jumps'' as a standard collisional kernel This result supports that the the center of mass frame corresponds to a ``no-jump frame'' previously proposed in Ref.~\cite{Chen:2015gta}. The details of computations incorporating background fields will be presented elsewhere.


\textit{Outlook}.---
{Our findings not only provide a solid footing for side-jump phenomena associated with Lorentz-symmetry properties of CKT from field theories}, but also raise some new issues. It turns out that a frame-independent distribution function can be introduced, but the trade-off is the modification of the spectral density and the corresponding CKT. However, in such a case, the CKT should be manifestly Lorentz invariant, whereas the derivation is nontrivial due to the presence of background fields, which will be pursued in the future. 

{Furthermore, the Wigner functions and the CKT influenced by collisions, can be applied to distinct systems with proper approximations of the self-energy, which may intrigue further studies for phenomenological interests. On the other hand, it is straightforward to include higher-order quantum corrections in our approach, which may reveal novel effects in a self-consistent fashion.}     
  
\textit{Acknowledgments}.---
We thank  M. Stephanov for useful discussions.
Y.H. was partially supported by JSPS KAKENHI Grants Numbers 15H03652, 16K17716 and the RIKEN interdisciplinary Theoretical Science (iTHES) project. S.P. is supported by JSPS post-doctoral fellowship for foreign researchers.
D.Y. was supported by the
RIKEN Foreign Postdoctoral Researcher program.


\begin{thebibliography}{33}%
\makeatletter
\providecommand \@ifxundefined [1]{%
 \@ifx{#1\undefined}
}%
\providecommand \@ifnum [1]{%
 \ifnum #1\expandafter \@firstoftwo
 \else \expandafter \@secondoftwo
 \fi
}%
\providecommand \@ifx [1]{%
 \ifx #1\expandafter \@firstoftwo
 \else \expandafter \@secondoftwo
 \fi
}%
\providecommand \natexlab [1]{#1}%
\providecommand \enquote  [1]{``#1''}%
\providecommand \bibnamefont  [1]{#1}%
\providecommand \bibfnamefont [1]{#1}%
\providecommand \citenamefont [1]{#1}%
\providecommand \href@noop [0]{\@secondoftwo}%
\providecommand \href [0]{\begingroup \@sanitize@url \@href}%
\providecommand \@href[1]{\@@startlink{#1}\@@href}%
\providecommand \@@href[1]{\endgroup#1\@@endlink}%
\providecommand \@sanitize@url [0]{\catcode `\\12\catcode `\$12\catcode
  `\&12\catcode `\#12\catcode `\^12\catcode `\_12\catcode `\%12\relax}%
\providecommand \@@startlink[1]{}%
\providecommand \@@endlink[0]{}%
\providecommand \url  [0]{\begingroup\@sanitize@url \@url }%
\providecommand \@url [1]{\endgroup\@href {#1}{\urlprefix }}%
\providecommand \urlprefix  [0]{URL }%
\providecommand \Eprint [0]{\href }%
\providecommand \doibase [0]{http://dx.doi.org/}%
\providecommand \selectlanguage [0]{\@gobble}%
\providecommand \bibinfo  [0]{\@secondoftwo}%
\providecommand \bibfield  [0]{\@secondoftwo}%
\providecommand \translation [1]{[#1]}%
\providecommand \BibitemOpen [0]{}%
\providecommand \bibitemStop [0]{}%
\providecommand \bibitemNoStop [0]{.\EOS\space}%
\providecommand \EOS [0]{\spacefactor3000\relax}%
\providecommand \BibitemShut  [1]{\csname bibitem#1\endcsname}%
\let\auto@bib@innerbib\@empty
\bibitem [{\citenamefont {Kharzeev}\ \emph {et~al.}(2008)\citenamefont
  {Kharzeev}, \citenamefont {McLerran},\ and\ \citenamefont
  {Warringa}}]{Kharzeev:2007jp}%
  \BibitemOpen
  \bibfield  {author} {\bibinfo {author} {\bibfnamefont {D.~E.}\ \bibnamefont
  {Kharzeev}}, \bibinfo {author} {\bibfnamefont {L.~D.}\ \bibnamefont
  {McLerran}}, \ and\ \bibinfo {author} {\bibfnamefont {H.~J.}\ \bibnamefont
  {Warringa}},\ }\href {\doibase 10.1016/j.nuclphysa.2008.02.298} {\bibfield
  {journal} {\bibinfo  {journal} {Nucl. Phys.}\ }\textbf {\bibinfo {volume}
  {A803}},\ \bibinfo {pages} {227} (\bibinfo {year} {2008})},\ \Eprint
  {http://arxiv.org/abs/0711.0950} {arXiv:0711.0950 [hep-ph]} \BibitemShut
  {NoStop}%
\bibitem [{\citenamefont {Fukushima}\ \emph {et~al.}(2008)\citenamefont
  {Fukushima}, \citenamefont {Kharzeev},\ and\ \citenamefont
  {Warringa}}]{Fukushima:2008xe}%
  \BibitemOpen
  \bibfield  {author} {\bibinfo {author} {\bibfnamefont {K.}~\bibnamefont
  {Fukushima}}, \bibinfo {author} {\bibfnamefont {D.~E.}\ \bibnamefont
  {Kharzeev}}, \ and\ \bibinfo {author} {\bibfnamefont {H.~J.}\ \bibnamefont
  {Warringa}},\ }\href {\doibase 10.1103/PhysRevD.78.074033} {\bibfield
  {journal} {\bibinfo  {journal} {Phys. Rev.}\ }\textbf {\bibinfo {volume}
  {D78}},\ \bibinfo {pages} {074033} (\bibinfo {year} {2008})},\ \Eprint
  {http://arxiv.org/abs/0808.3382} {arXiv:0808.3382 [hep-ph]} \BibitemShut
  {NoStop}%
\bibitem [{\citenamefont {Kharzeev}\ and\ \citenamefont
  {Warringa}(2009)}]{Kharzeev:2009pj}%
  \BibitemOpen
  \bibfield  {author} {\bibinfo {author} {\bibfnamefont {D.~E.}\ \bibnamefont
  {Kharzeev}}\ and\ \bibinfo {author} {\bibfnamefont {H.~J.}\ \bibnamefont
  {Warringa}},\ }\href {\doibase 10.1103/PhysRevD.80.034028} {\bibfield
  {journal} {\bibinfo  {journal} {Phys. Rev.}\ }\textbf {\bibinfo {volume}
  {D80}},\ \bibinfo {pages} {034028} (\bibinfo {year} {2009})},\ \Eprint
  {http://arxiv.org/abs/0907.5007} {arXiv:0907.5007 [hep-ph]} \BibitemShut
  {NoStop}%
\bibitem [{\citenamefont {Son}\ and\ \citenamefont
  {Surowka}(2009)}]{Son:2009tf}%
  \BibitemOpen
  \bibfield  {author} {\bibinfo {author} {\bibfnamefont {D.~T.}\ \bibnamefont
  {Son}}\ and\ \bibinfo {author} {\bibfnamefont {P.}~\bibnamefont {Surowka}},\
  }\href {\doibase 10.1103/PhysRevLett.103.191601} {\bibfield  {journal}
  {\bibinfo  {journal} {Phys. Rev. Lett.}\ }\textbf {\bibinfo {volume} {103}},\
  \bibinfo {pages} {191601} (\bibinfo {year} {2009})},\ \Eprint
  {http://arxiv.org/abs/0906.5044} {arXiv:0906.5044 [hep-th]} \BibitemShut
  {NoStop}%
\bibitem [{\citenamefont {Sadofyev}\ and\ \citenamefont
  {Isachenkov}(2011)}]{Sadofyev:2010pr}%
  \BibitemOpen
  \bibfield  {author} {\bibinfo {author} {\bibfnamefont {A.~V.}\ \bibnamefont
  {Sadofyev}}\ and\ \bibinfo {author} {\bibfnamefont {M.~V.}\ \bibnamefont
  {Isachenkov}},\ }\href {\doibase 10.1016/j.physletb.2011.02.041} {\bibfield
  {journal} {\bibinfo  {journal} {Phys. Lett.}\ }\textbf {\bibinfo {volume}
  {B697}},\ \bibinfo {pages} {404} (\bibinfo {year} {2011})},\ \Eprint
  {http://arxiv.org/abs/1010.1550} {arXiv:1010.1550 [hep-th]} \BibitemShut
  {NoStop}%
\bibitem [{\citenamefont {Pu}\ \emph {et~al.}(2011)\citenamefont {Pu},
  \citenamefont {Gao},\ and\ \citenamefont {Wang}}]{Pu:2010as}%
  \BibitemOpen
  \bibfield  {author} {\bibinfo {author} {\bibfnamefont {S.}~\bibnamefont
  {Pu}}, \bibinfo {author} {\bibfnamefont {J.-h.}\ \bibnamefont {Gao}}, \ and\
  \bibinfo {author} {\bibfnamefont {Q.}~\bibnamefont {Wang}},\ }\href {\doibase
  10.1103/PhysRevD.83.094017} {\bibfield  {journal} {\bibinfo  {journal} {Phys.
  Rev.}\ }\textbf {\bibinfo {volume} {D83}},\ \bibinfo {pages} {094017}
  (\bibinfo {year} {2011})},\ \Eprint {http://arxiv.org/abs/1008.2418}
  {arXiv:1008.2418 [nucl-th]} \BibitemShut {NoStop}%
\bibitem [{\citenamefont {Kharzeev}\ and\ \citenamefont
  {Yee}(2011)}]{Kharzeev:2011ds}%
  \BibitemOpen
  \bibfield  {author} {\bibinfo {author} {\bibfnamefont {D.~E.}\ \bibnamefont
  {Kharzeev}}\ and\ \bibinfo {author} {\bibfnamefont {H.-U.}\ \bibnamefont
  {Yee}},\ }\href {\doibase 10.1103/PhysRevD.84.045025} {\bibfield  {journal}
  {\bibinfo  {journal} {Phys. Rev.}\ }\textbf {\bibinfo {volume} {D84}},\
  \bibinfo {pages} {045025} (\bibinfo {year} {2011})},\ \Eprint
  {http://arxiv.org/abs/1105.6360} {arXiv:1105.6360 [hep-th]} \BibitemShut
  {NoStop}%
\bibitem [{\citenamefont {Abramczyk}\ \emph {et~al.}(2009)\citenamefont
  {Abramczyk}, \citenamefont {Blum}, \citenamefont {Petropoulos},\ and\
  \citenamefont {Zhou}}]{Abramczyk:2009gb}%
  \BibitemOpen
  \bibfield  {author} {\bibinfo {author} {\bibfnamefont {M.}~\bibnamefont
  {Abramczyk}}, \bibinfo {author} {\bibfnamefont {T.}~\bibnamefont {Blum}},
  \bibinfo {author} {\bibfnamefont {G.}~\bibnamefont {Petropoulos}}, \ and\
  \bibinfo {author} {\bibfnamefont {R.}~\bibnamefont {Zhou}},\ }\href@noop {}
  {\bibfield  {journal} {\bibinfo  {journal} {PoS}\ }\textbf {\bibinfo {volume}
  {LAT2009}},\ \bibinfo {pages} {181} (\bibinfo {year} {2009})},\ \Eprint
  {http://arxiv.org/abs/0911.1348} {arXiv:0911.1348 [hep-lat]} \BibitemShut
  {NoStop}%
\bibitem [{\citenamefont {Buividovich}\ \emph
  {et~al.}(2009{\natexlab{a}})\citenamefont {Buividovich}, \citenamefont
  {Chernodub}, \citenamefont {Luschevskaya},\ and\ \citenamefont
  {Polikarpov}}]{Buividovich:2009wi}%
  \BibitemOpen
  \bibfield  {author} {\bibinfo {author} {\bibfnamefont {P.}~\bibnamefont
  {Buividovich}}, \bibinfo {author} {\bibfnamefont {M.}~\bibnamefont
  {Chernodub}}, \bibinfo {author} {\bibfnamefont {E.}~\bibnamefont
  {Luschevskaya}}, \ and\ \bibinfo {author} {\bibfnamefont {M.}~\bibnamefont
  {Polikarpov}},\ }\href {\doibase 10.1103/PhysRevD.80.054503} {\bibfield
  {journal} {\bibinfo  {journal} {Phys. Rev.}\ }\textbf {\bibinfo {volume}
  {D80}},\ \bibinfo {pages} {054503} (\bibinfo {year} {2009}{\natexlab{a}})},\
  \Eprint {http://arxiv.org/abs/0907.0494} {arXiv:0907.0494 [hep-lat]}
  \BibitemShut {NoStop}%
\bibitem [{\citenamefont {Buividovich}\ \emph
  {et~al.}(2009{\natexlab{b}})\citenamefont {Buividovich}, \citenamefont
  {Luschevskaya}, \citenamefont {Polikarpov},\ and\ \citenamefont
  {Chernodub}}]{Buividovich:2009zzb}%
  \BibitemOpen
  \bibfield  {author} {\bibinfo {author} {\bibfnamefont {P.}~\bibnamefont
  {Buividovich}}, \bibinfo {author} {\bibfnamefont {E.}~\bibnamefont
  {Luschevskaya}}, \bibinfo {author} {\bibfnamefont {M.}~\bibnamefont
  {Polikarpov}}, \ and\ \bibinfo {author} {\bibfnamefont {M.}~\bibnamefont
  {Chernodub}},\ }\href {\doibase 10.1134/S0021364009180027} {\bibfield
  {journal} {\bibinfo  {journal} {JETP Lett.}\ }\textbf {\bibinfo {volume}
  {90}},\ \bibinfo {pages} {412} (\bibinfo {year}
  {2009}{\natexlab{b}})}\BibitemShut {NoStop}%
\bibitem [{\citenamefont {Buividovich}\ \emph {et~al.}(2010)\citenamefont
  {Buividovich}, \citenamefont {Chernodub}, \citenamefont {Kharzeev},
  \citenamefont {Kalaydzhyan}, \citenamefont {Luschevskaya} \emph
  {et~al.}}]{Buividovich:2010tn}%
  \BibitemOpen
  \bibfield  {author} {\bibinfo {author} {\bibfnamefont {P.}~\bibnamefont
  {Buividovich}}, \bibinfo {author} {\bibfnamefont {M.}~\bibnamefont
  {Chernodub}}, \bibinfo {author} {\bibfnamefont {D.}~\bibnamefont {Kharzeev}},
  \bibinfo {author} {\bibfnamefont {T.}~\bibnamefont {Kalaydzhyan}}, \bibinfo
  {author} {\bibfnamefont {E.}~\bibnamefont {Luschevskaya}},  \emph {et~al.},\
  }\href {\doibase 10.1103/PhysRevLett.105.132001} {\bibfield  {journal}
  {\bibinfo  {journal} {Phys. Rev. Lett.}\ }\textbf {\bibinfo {volume} {105}},\
  \bibinfo {pages} {132001} (\bibinfo {year} {2010})},\ \Eprint
  {http://arxiv.org/abs/1003.2180} {arXiv:1003.2180 [hep-lat]} \BibitemShut
  {NoStop}%
\bibitem [{\citenamefont {Yamamoto}(2011)}]{Yamamoto:2011gk}%
  \BibitemOpen
  \bibfield  {author} {\bibinfo {author} {\bibfnamefont {A.}~\bibnamefont
  {Yamamoto}},\ }\href {\doibase 10.1103/PhysRevLett.107.031601} {\bibfield
  {journal} {\bibinfo  {journal} {Phys. Rev. Lett.}\ }\textbf {\bibinfo
  {volume} {107}},\ \bibinfo {pages} {031601} (\bibinfo {year} {2011})},\
  \Eprint {http://arxiv.org/abs/1105.0385} {arXiv:1105.0385 [hep-lat]}
  \BibitemShut {NoStop}%
\bibitem [{\citenamefont {Erdmenger}\ \emph {et~al.}(2009)\citenamefont
  {Erdmenger}, \citenamefont {Haack}, \citenamefont {Kaminski},\ and\
  \citenamefont {Yarom}}]{Erdmenger2009}%
  \BibitemOpen
  \bibfield  {author} {\bibinfo {author} {\bibfnamefont {J.}~\bibnamefont
  {Erdmenger}}, \bibinfo {author} {\bibfnamefont {M.}~\bibnamefont {Haack}},
  \bibinfo {author} {\bibfnamefont {M.}~\bibnamefont {Kaminski}}, \ and\
  \bibinfo {author} {\bibfnamefont {A.}~\bibnamefont {Yarom}},\ }\href
  {\doibase 10.1088/1126-6708/2009/01/055} {\bibfield  {journal} {\bibinfo
  {journal} {JHEP}\ }\textbf {\bibinfo {volume} {01}},\ \bibinfo {pages} {055}
  (\bibinfo {year} {2009})},\ \Eprint {http://arxiv.org/abs/0809.2488}
  {arXiv:0809.2488 [hep-th]} \BibitemShut {NoStop}%
\bibitem [{\citenamefont {Torabian}\ and\ \citenamefont
  {Yee}(2009)}]{Torabian2009a}%
  \BibitemOpen
  \bibfield  {author} {\bibinfo {author} {\bibfnamefont {M.}~\bibnamefont
  {Torabian}}\ and\ \bibinfo {author} {\bibfnamefont {H.-U.}\ \bibnamefont
  {Yee}},\ }\href {\doibase 10.1088/1126-6708/2009/08/020} {\bibfield
  {journal} {\bibinfo  {journal} {JHEP}\ }\textbf {\bibinfo {volume} {08}},\
  \bibinfo {pages} {020} (\bibinfo {year} {2009})},\ \Eprint
  {http://arxiv.org/abs/0903.4894} {arXiv:0903.4894 [hep-th]} \BibitemShut
  {NoStop}%
\bibitem [{\citenamefont {Banerjee}\ \emph {et~al.}(2011)\citenamefont
  {Banerjee}, \citenamefont {Bhattacharya}, \citenamefont {Bhattacharyya},
  \citenamefont {Dutta}, \citenamefont {Loganayagam} \emph
  {et~al.}}]{Banerjee2011}%
  \BibitemOpen
  \bibfield  {author} {\bibinfo {author} {\bibfnamefont {N.}~\bibnamefont
  {Banerjee}}, \bibinfo {author} {\bibfnamefont {J.}~\bibnamefont
  {Bhattacharya}}, \bibinfo {author} {\bibfnamefont {S.}~\bibnamefont
  {Bhattacharyya}}, \bibinfo {author} {\bibfnamefont {S.}~\bibnamefont
  {Dutta}}, \bibinfo {author} {\bibfnamefont {R.}~\bibnamefont {Loganayagam}},
  \emph {et~al.},\ }\href {\doibase 10.1007/JHEP01(2011)094} {\bibfield
  {journal} {\bibinfo  {journal} {JHEP}\ }\textbf {\bibinfo {volume} {1101}},\
  \bibinfo {pages} {094} (\bibinfo {year} {2011})},\ \Eprint
  {http://arxiv.org/abs/0809.2596} {arXiv:0809.2596 [hep-th]} \BibitemShut
  {NoStop}%
\bibitem [{\citenamefont {Kharzeev}\ \emph {et~al.}(2016)\citenamefont
  {Kharzeev}, \citenamefont {Liao}, \citenamefont {Voloshin},\ and\
  \citenamefont {Wang}}]{Kharzeev:2015znc}%
  \BibitemOpen
  \bibfield  {author} {\bibinfo {author} {\bibfnamefont {D.~E.}\ \bibnamefont
  {Kharzeev}}, \bibinfo {author} {\bibfnamefont {J.}~\bibnamefont {Liao}},
  \bibinfo {author} {\bibfnamefont {S.~A.}\ \bibnamefont {Voloshin}}, \ and\
  \bibinfo {author} {\bibfnamefont {G.}~\bibnamefont {Wang}},\ }\href {\doibase
  10.1016/j.ppnp.2016.01.001} {\bibfield  {journal} {\bibinfo  {journal} {Prog.
  Part. Nucl. Phys.}\ }\textbf {\bibinfo {volume} {88}},\ \bibinfo {pages} {1}
  (\bibinfo {year} {2016})},\ \Eprint {http://arxiv.org/abs/1511.04050}
  {arXiv:1511.04050 [hep-ph]} \BibitemShut {NoStop}%
\bibitem [{\citenamefont {Li}\ \emph {et~al.}(2016)\citenamefont {Li},
  \citenamefont {Kharzeev}, \citenamefont {Zhang}, \citenamefont {Huang},
  \citenamefont {Pletikosic}, \citenamefont {Fedorov}, \citenamefont {Zhong},
  \citenamefont {Schneeloch}, \citenamefont {Gu},\ and\ \citenamefont
  {Valla}}]{Li:2014bha}%
  \BibitemOpen
  \bibfield  {author} {\bibinfo {author} {\bibfnamefont {Q.}~\bibnamefont
  {Li}}, \bibinfo {author} {\bibfnamefont {D.~E.}\ \bibnamefont {Kharzeev}},
  \bibinfo {author} {\bibfnamefont {C.}~\bibnamefont {Zhang}}, \bibinfo
  {author} {\bibfnamefont {Y.}~\bibnamefont {Huang}}, \bibinfo {author}
  {\bibfnamefont {I.}~\bibnamefont {Pletikosic}}, \bibinfo {author}
  {\bibfnamefont {A.~V.}\ \bibnamefont {Fedorov}}, \bibinfo {author}
  {\bibfnamefont {R.~D.}\ \bibnamefont {Zhong}}, \bibinfo {author}
  {\bibfnamefont {J.~A.}\ \bibnamefont {Schneeloch}}, \bibinfo {author}
  {\bibfnamefont {G.~D.}\ \bibnamefont {Gu}}, \ and\ \bibinfo {author}
  {\bibfnamefont {T.}~\bibnamefont {Valla}},\ }\href {\doibase
  10.1038/nphys3648} {\bibfield  {journal} {\bibinfo  {journal} {Nature Phys.}\
  }\textbf {\bibinfo {volume} {12}},\ \bibinfo {pages} {550} (\bibinfo {year}
  {2016})},\ \Eprint {http://arxiv.org/abs/1412.6543} {arXiv:1412.6543
  [cond-mat.str-el]} \BibitemShut {NoStop}%
\bibitem [{\citenamefont {Stephanov}\ and\ \citenamefont
  {Yin}(2012)}]{Stephanov:2012ki}%
  \BibitemOpen
  \bibfield  {author} {\bibinfo {author} {\bibfnamefont {M.}~\bibnamefont
  {Stephanov}}\ and\ \bibinfo {author} {\bibfnamefont {Y.}~\bibnamefont
  {Yin}},\ }\href {\doibase 10.1103/PhysRevLett.109.162001} {\bibfield
  {journal} {\bibinfo  {journal} {Phys. Rev. Lett.}\ }\textbf {\bibinfo
  {volume} {109}},\ \bibinfo {pages} {162001} (\bibinfo {year} {2012})},\
  \Eprint {http://arxiv.org/abs/1207.0747} {arXiv:1207.0747 [hep-th]}
  \BibitemShut {NoStop}%
\bibitem [{\citenamefont {Son}\ and\ \citenamefont
  {Yamamoto}(2012)}]{Son:2012wh}%
  \BibitemOpen
  \bibfield  {author} {\bibinfo {author} {\bibfnamefont {D.~T.}\ \bibnamefont
  {Son}}\ and\ \bibinfo {author} {\bibfnamefont {N.}~\bibnamefont {Yamamoto}},\
  }\href {\doibase 10.1103/PhysRevLett.109.181602} {\bibfield  {journal}
  {\bibinfo  {journal} {Phys. Rev. Lett.}\ }\textbf {\bibinfo {volume} {109}},\
  \bibinfo {pages} {181602} (\bibinfo {year} {2012})},\ \Eprint
  {http://arxiv.org/abs/1203.2697} {arXiv:1203.2697 [cond-mat.mes-hall]}
  \BibitemShut {NoStop}%
\bibitem [{\citenamefont {Gao}\ \emph {et~al.}(2012)\citenamefont {Gao},
  \citenamefont {Liang}, \citenamefont {Pu}, \citenamefont {Wang},\ and\
  \citenamefont {Wang}}]{Gao:2012ix}%
  \BibitemOpen
  \bibfield  {author} {\bibinfo {author} {\bibfnamefont {J.-H.}\ \bibnamefont
  {Gao}}, \bibinfo {author} {\bibfnamefont {Z.-T.}\ \bibnamefont {Liang}},
  \bibinfo {author} {\bibfnamefont {S.}~\bibnamefont {Pu}}, \bibinfo {author}
  {\bibfnamefont {Q.}~\bibnamefont {Wang}}, \ and\ \bibinfo {author}
  {\bibfnamefont {X.-N.}\ \bibnamefont {Wang}},\ }\href {\doibase
  10.1103/PhysRevLett.109.232301} {\bibfield  {journal} {\bibinfo  {journal}
  {Phys. Rev. Lett.}\ }\textbf {\bibinfo {volume} {109}},\ \bibinfo {pages}
  {232301} (\bibinfo {year} {2012})},\ \Eprint {http://arxiv.org/abs/1203.0725}
  {arXiv:1203.0725 [hep-ph]} \BibitemShut {NoStop}%
\bibitem [{\citenamefont {Chen}\ \emph {et~al.}(2013)\citenamefont {Chen},
  \citenamefont {Pu}, \citenamefont {Wang},\ and\ \citenamefont
  {Wang}}]{Chen:2012ca}%
  \BibitemOpen
  \bibfield  {author} {\bibinfo {author} {\bibfnamefont {J.-W.}\ \bibnamefont
  {Chen}}, \bibinfo {author} {\bibfnamefont {S.}~\bibnamefont {Pu}}, \bibinfo
  {author} {\bibfnamefont {Q.}~\bibnamefont {Wang}}, \ and\ \bibinfo {author}
  {\bibfnamefont {X.-N.}\ \bibnamefont {Wang}},\ }\href {\doibase
  10.1103/PhysRevLett.110.262301} {\bibfield  {journal} {\bibinfo  {journal}
  {Phys. Rev. Lett.}\ }\textbf {\bibinfo {volume} {110}},\ \bibinfo {pages}
  {262301} (\bibinfo {year} {2013})},\ \Eprint {http://arxiv.org/abs/1210.8312}
  {arXiv:1210.8312 [hep-th]} \BibitemShut {NoStop}%
\bibitem [{\citenamefont {Berry}(1984)}]{Berry1984}%
  \BibitemOpen
  \bibfield  {author} {\bibinfo {author} {\bibfnamefont {M.~V.}\ \bibnamefont
  {Berry}},\ }\href {\doibase 10.1098/rspa.1984.0023} {\bibfield  {journal}
  {\bibinfo  {journal} {Proc. Roy. Soc. Lond.}\ }\textbf {\bibinfo {volume}
  {A392}},\ \bibinfo {pages} {45} (\bibinfo {year} {1984})}\BibitemShut
  {NoStop}%
\bibitem [{\citenamefont {Chen}\ \emph
  {et~al.}(2014{\natexlab{a}})\citenamefont {Chen}, \citenamefont {Pang},
  \citenamefont {Pu},\ and\ \citenamefont {Wang}}]{Chen:2013iga}%
  \BibitemOpen
  \bibfield  {author} {\bibinfo {author} {\bibfnamefont {J.-W.}\ \bibnamefont
  {Chen}}, \bibinfo {author} {\bibfnamefont {J.-y.}\ \bibnamefont {Pang}},
  \bibinfo {author} {\bibfnamefont {S.}~\bibnamefont {Pu}}, \ and\ \bibinfo
  {author} {\bibfnamefont {Q.}~\bibnamefont {Wang}},\ }\href {\doibase
  10.1103/PhysRevD.89.094003} {\bibfield  {journal} {\bibinfo  {journal} {Phys.
  Rev.}\ }\textbf {\bibinfo {volume} {D89}},\ \bibinfo {pages} {094003}
  (\bibinfo {year} {2014}{\natexlab{a}})},\ \Eprint
  {http://arxiv.org/abs/1312.2032} {arXiv:1312.2032 [hep-th]} \BibitemShut
  {NoStop}%
\bibitem [{\citenamefont {Son}\ and\ \citenamefont
  {Yamamoto}(2013)}]{Son:2012zy}%
  \BibitemOpen
  \bibfield  {author} {\bibinfo {author} {\bibfnamefont {D.~T.}\ \bibnamefont
  {Son}}\ and\ \bibinfo {author} {\bibfnamefont {N.}~\bibnamefont {Yamamoto}},\
  }\href {\doibase 10.1103/PhysRevD.87.085016} {\bibfield  {journal} {\bibinfo
  {journal} {Phys. Rev.}\ }\textbf {\bibinfo {volume} {D87}},\ \bibinfo {pages}
  {085016} (\bibinfo {year} {2013})},\ \Eprint {http://arxiv.org/abs/1210.8158}
  {arXiv:1210.8158 [hep-th]} \BibitemShut {NoStop}%
\bibitem [{\citenamefont {Manuel}\ and\ \citenamefont
  {Torres-Rincon}(2014)}]{Manuel:2014dza}%
  \BibitemOpen
  \bibfield  {author} {\bibinfo {author} {\bibfnamefont {C.}~\bibnamefont
  {Manuel}}\ and\ \bibinfo {author} {\bibfnamefont {J.~M.}\ \bibnamefont
  {Torres-Rincon}},\ }\href {\doibase 10.1103/PhysRevD.90.076007} {\bibfield
  {journal} {\bibinfo  {journal} {Phys. Rev.}\ }\textbf {\bibinfo {volume}
  {D90}},\ \bibinfo {pages} {076007} (\bibinfo {year} {2014})},\ \Eprint
  {http://arxiv.org/abs/1404.6409} {arXiv:1404.6409 [hep-ph]} \BibitemShut
  {NoStop}%
\bibitem [{\citenamefont {Chen}\ \emph
  {et~al.}(2014{\natexlab{b}})\citenamefont {Chen}, \citenamefont {Son},
  \citenamefont {Stephanov}, \citenamefont {Yee},\ and\ \citenamefont
  {Yin}}]{Chen:2014cla}%
  \BibitemOpen
  \bibfield  {author} {\bibinfo {author} {\bibfnamefont {J.-Y.}\ \bibnamefont
  {Chen}}, \bibinfo {author} {\bibfnamefont {D.~T.}\ \bibnamefont {Son}},
  \bibinfo {author} {\bibfnamefont {M.~A.}\ \bibnamefont {Stephanov}}, \bibinfo
  {author} {\bibfnamefont {H.-U.}\ \bibnamefont {Yee}}, \ and\ \bibinfo
  {author} {\bibfnamefont {Y.}~\bibnamefont {Yin}},\ }\href {\doibase
  10.1103/PhysRevLett.113.182302} {\bibfield  {journal} {\bibinfo  {journal}
  {Phys. Rev. Lett.}\ }\textbf {\bibinfo {volume} {113}},\ \bibinfo {pages}
  {182302} (\bibinfo {year} {2014}{\natexlab{b}})},\ \Eprint
  {http://arxiv.org/abs/1404.5963} {arXiv:1404.5963 [hep-th]} \BibitemShut
  {NoStop}%
\bibitem [{\citenamefont {Chen}\ \emph {et~al.}(2015)\citenamefont {Chen},
  \citenamefont {Son},\ and\ \citenamefont {Stephanov}}]{Chen:2015gta}%
  \BibitemOpen
  \bibfield  {author} {\bibinfo {author} {\bibfnamefont {J.-Y.}\ \bibnamefont
  {Chen}}, \bibinfo {author} {\bibfnamefont {D.~T.}\ \bibnamefont {Son}}, \
  and\ \bibinfo {author} {\bibfnamefont {M.~A.}\ \bibnamefont {Stephanov}},\
  }\href {\doibase 10.1103/PhysRevLett.115.021601} {\bibfield  {journal}
  {\bibinfo  {journal} {Phys. Rev. Lett.}\ }\textbf {\bibinfo {volume} {115}},\
  \bibinfo {pages} {021601} (\bibinfo {year} {2015})},\ \Eprint
  {http://arxiv.org/abs/1502.06966} {arXiv:1502.06966 [hep-th]} \BibitemShut
  {NoStop}%
\bibitem [{\citenamefont {Elze}\ \emph {et~al.}(1986)\citenamefont {Elze},
  \citenamefont {Gyulassy},\ and\ \citenamefont {Vasak}}]{Elze:1986qd}%
  \BibitemOpen
  \bibfield  {author} {\bibinfo {author} {\bibfnamefont {H.~T.}\ \bibnamefont
  {Elze}}, \bibinfo {author} {\bibfnamefont {M.}~\bibnamefont {Gyulassy}}, \
  and\ \bibinfo {author} {\bibfnamefont {D.}~\bibnamefont {Vasak}},\ }\href
  {\doibase 10.1016/0550-3213(86)90072-6} {\bibfield  {journal} {\bibinfo
  {journal} {Nucl. Phys.}\ }\textbf {\bibinfo {volume} {B276}},\ \bibinfo
  {pages} {706} (\bibinfo {year} {1986})}\BibitemShut {NoStop}%
\bibitem [{Note1()}]{Note1}%
  \BibitemOpen
  \bibinfo {note} {Note that one needs to use $\delta (q^2)q^{\mu }\Delta _{\mu
  }f=0$ to substitute $\Delta _0f$ with $q_0^{-1}{\protect \bf q}\cdot
  {\protect \bf \Delta }f$ in order to show the solution satisfies the first
  equation in Eq.~(\ref {traceless_part}). Also, one can show that Eq.~(\ref
  {trace_sol}) satisfies Eqs.~(\ref {trace_part}) and (\ref {traceless_part})
  for constant $f$.}\BibitemShut {Stop}%
\bibitem [{Note2()}]{Note2}%
  \BibitemOpen
  \bibinfo {note} {Note that the gradient expansion implicitly introduces an
  infrared cutoff for $|{\protect \bf q}|$, which has to be larger than the
  scale of gradient or background fields.}\BibitemShut {Stop}%
\bibitem [{Note3()}]{Note3}%
  \BibitemOpen
  \bibinfo {note} {One can in fact consider the frame dependence of Eq.~(\ref
  {traceless_part}) and check that the following solution indeed satisfies both
  equations therein.}\BibitemShut {Stop}%
\bibitem [{\citenamefont {Weinberg}(1995)}]{Weinberg_vol1}%
  \BibitemOpen
  \bibfield  {author} {\bibinfo {author} {\bibfnamefont {S.}~\bibnamefont
  {Weinberg}},\ }\href@noop {} {\emph {\bibinfo {title} {The Quantum Theory of
  Fields, Volume I}}}\ (\bibinfo  {publisher} {Cambridge University Press},\
  \bibinfo {year} {1995})\BibitemShut {NoStop}%
\bibitem [{\citenamefont {Blaizot}\ and\ \citenamefont
  {Iancu}(2002)}]{Blaizot:2001nr}%
  \BibitemOpen
  \bibfield  {author} {\bibinfo {author} {\bibfnamefont {J.-P.}\ \bibnamefont
  {Blaizot}}\ and\ \bibinfo {author} {\bibfnamefont {E.}~\bibnamefont
  {Iancu}},\ }\href {\doibase 10.1016/S0370-1573(01)00061-8} {\bibfield
  {journal} {\bibinfo  {journal} {Phys. Rept.}\ }\textbf {\bibinfo {volume}
  {359}},\ \bibinfo {pages} {355} (\bibinfo {year} {2002})},\ \Eprint
  {http://arxiv.org/abs/hep-ph/0101103} {arXiv:hep-ph/0101103 [hep-ph]}
  \BibitemShut {NoStop}%
\end{thebibliography}
\end{document}